\documentclass[review]{elsarticle}
\usepackage{graphicx}
\usepackage{lineno,hyperref}
\usepackage{dcolumn}
\usepackage{bm}
\usepackage{epsfig}
\usepackage{caption}
\usepackage[english]{babel}
\usepackage[english]{babel}

\modulolinenumbers[5]

\journal{???}

\begin{document}

\begin{frontmatter}

\title{Numerical path integral solution to strong Coulomb correlation in one dimensional Hooke's atom}

\author{Ilkka Ruokosenm\"aki, Hosein Gholizade, Ilkka Kyl\"anp\"a\"a and Tapio T.~Rantala}
\address{Department of Physics, Tampere University of Technology, Finland}

\begin{abstract}
We present a new approach based on real time domain Feynman path
integrals (RTPI) for electronic structure calculations and quantum
dynamics, which includes correlations between particles exactly but within
the numerical accuracy. We demonstrate that incoherent propagation by
keeping the wave function real is a novel method for finding and
simulation of the ground state, similar to Diffusion Monte Carlo (DMC)
method, but introducing new useful tools lacking in DMC.  We use 1D
Hooke's atom, a two-electron system with very strong correlation, as
our test case, which we solve with incoherent RTPI (iRTPI) and compare against
DMC. This system provides an excellent test case due to exact solutions
for some confinements and because in 1D the Coulomb singularity is
stronger than in two or three dimensional space. The use of Monte
Carlo grid is shown to be efficient for which we determine useful
numerical parameters. Furthermore, we discuss another novel approach
achieved by combining the strengths of iRTPI and DMC.
We also show usefulness of the perturbation theory for analytical
approximates in case of strong confinements.
\end{abstract}


\end{frontmatter}

\pagebreak


\section{Introduction}

Feynman path integral (PI) approach offers an intuitive description of quantum mechanics \cite{feynman,feynman2}, where classical mechanics emerges transparently from disappearing wave nature of particles along with vanishing Planck constant.  Therefore, it is robust in numerical calculations in cases close to classical ones, like molecular quantum dynamics in real time \cite{makri}, but becomes more challenging and laborious for states of electrons, where the wave nature plays larger role.
Furthermore, the PI presentation of stationary states also involves full time-dependent quantum dynamics, in contrast with the conventional solution of the time-dependent Schr\"odinger equation, where time evolution appears as simple change of the wave function phase, only.

We have already demonstrated that numerical solutions to stationary states and quantum dynamics of single electrons in one dimensional potentials can be reliably found, both in regular and Monte Carlo grids, by using real time path integral (RTPI) propagation \cite{paper1}.  We have also assessed the usefulness and accuracy of the Trotter kernel as compared to the exact kernels and pointed out the advantages of the Monte Carlo grid in avoiding spurious interference effects.  For search and evaluation of the single particle eigenstates we found a novel approach based on the \emph{incoherent propagation} \cite{paper1}, {\it i.e.}, collapsing the wave function to its real component after each short time step.  This is the starting point of the present study.

RTPI approach can be expected to show most of its proficiency in simulation of many-electron systems, where correlation phenomena turn out to be in major role -- the same way and partly for same reasons as it has been found to be with the more conventional path integral Monte Carlo (PIMC), simulation of the imaginary time propagation \cite{ceperley,kylanpaa,kylanpaaP,militzer}.  It may be pertinent to point out, that while PIMC simulation yields the finite temperature equilibrium description of the system of quantum particles, RTPI simulation finds the zero-Kelvin real time quantum dynamics. Furthermore, RTPI can also be used to find and simulate the eigenstates, as indicated above.  Thus, for finding and simulation of the ground state, RTPI can be compared to the diffusion Monte Carlo (DMC) simulation \cite{DMCbook}.  Thus, combination of these two can be expected to offer novel features, which turns out to be the case.

To assess the performance of incoherent RTPI as compared with DMC we choose the Hooke's atom in one dimension as the test bench, presenting a case of an extremely strong correlation.

Three dimensional Hooke's atom is a helium-like system of two electrons with Coulomb repulsion, where electron--nucleus attraction is replaced by a confining parabolic or harmonic potential.  It is one of the few non-trivial systems with exact solutions for certain strengths of confinement (harmonic force constant) \cite{taut}, and therefore, it is a good test case for our new approach.  As shown below, separation of the three dimensional problem in relative coordinates yields two problems, one of which is the one dimensional Hooke's atom once the angular momentum degrees of freedom are taken out. In one dimension, the Coulomb repulsion is strong enough to split the space to two independent domains defined by exchange of the electrons\cite{oseguera,takasu,Girardeau,Volosniev}.

In this paper we will demonstrate the novel incoherent RTPI in finding the ground state of one dimensional Hooke's atom by using a Monte Carlo grid.  We also analyze performance of the simulation by comparison with DMC simulation, and furthermore, discuss another new idea to combine the strengths of incoherent RTPI and DMC.  Accuracy of the numerical approaches is analyzed by using analytical solutions and those from perturbation theory (PT), where relevant.

\goodbreak

\section{Ground state}

Finding or simulation of the ground state is perhaps the most general problem to work out in dealing with quantum systems.  Here, we present our novel approach to this based on the incoherent real time propagation \cite{paper1} using the path integral formalism.  First however, we briefly present the well known diffusion Monte Carlo (DMC) method \cite{DMCbook} using imaginary time propagation, to be used as a reference.  These both are numerical methods and the former one in its robust form also using Monte Carlo technique.  For the specified test case, one dimensional Hooke's atom, we also compare with the analytical solutions, where available, and approximate solutions otherwise.

To keep notations simple, we use the atomic units, where \( m_e \) = \( e \) =  \( \hbar \) = \( 4\pi\epsilon_0\) = 1 throughout the paper, unless otherwise stated.

\subsection{Imaginary time propagation: DMC}

The time-dependent Schr\" odinger wave equation for the many-body wave function \( \psi(x,t) \) is
\begin{equation}  \label{SE1}  
{\rm i} \frac{\partial \psi(x,t)}{ \partial t} = (H - E_T) \psi(x,t),
\end{equation}
where \( H\) is the hamiltonian, \( x\) stands for all coordinates of particles in one or more spatial dimensions and \( E_T \) is an arbitrary reference energy or shift of zero level. Now, by replacing the real time \(t\) by imaginary time \( \tau = it \), this becomes
\begin{equation}  \label{SE2}  
- \frac{\partial \psi(x,\tau)}{\partial \tau} = (H - E_T) \psi(x,\tau),
\end{equation}
which is of the form of a diffusion equation.  Its solutions can be expressed in terms of eigenfunctions \( \phi_n(x) \) of the hamiltonian as
\begin{equation} \label{SO1} 
\psi(x,\tau) = \sum_{n=0}^{\infty} C_n \phi_n(x) \exp[-(E_n - E_T)\tau].
\end{equation}

As Monte Carlo methods are useful for evaluation of integrals, the differential equation is transformed into an integral equation. This is done by using Green's function formalism \cite{DMCbook} and we seek the solution of the form
\begin{equation} \label{SO5} 
\psi(x_b,\tau_b) = \int_a G(x_b,\tau_b;x_a,\tau_a)\psi(x_a,\tau_a)dx_a,
\end{equation}
where \( G(x_b,\tau_b;x_a,\tau_a) \) is the Green's function of the system, the position space representation of the time evolution operator \( \exp[-(H-E_T) (\tau_b - \tau_a) ] \).

The exact analytical form of the Green's function is rarely known, and therefore, it needs to be approximated.  Use of the so called short time approximation \cite{DMCbook} to separate the kinetic and potential energy contributions, \( T \) and \( V \), gives
\begin{equation} \label{shorttime} 
\exp[-(H-E_T)\Delta\tau] = \exp[-(T+V-E_T)\Delta\tau] \approx \exp[-T \Delta\tau] \exp[-(V-E_T) \Delta\tau].
\end{equation}
Since $T$ and $V$ do not commute, in general, this approximation is exact only in the limit \( \Delta\tau \rightarrow 0 \) but accurate for small \( \Delta\tau \) for potentials bound from below \cite{DMCbook}.

The Green's function can be separated into two parts,  kinetic and potential (or diffusion and branching),
\begin{equation} \label{shorttimeG} 
G(x_b,\tau_b;x,\tau_a) \approx G_{\rm diff}(x_b,\tau_b;x,\tau_a)G_{B}(x_b,\tau_b;x,\tau_a).
\end{equation}
As this Green's function satisfies the imaginary time Schr\"odinger equation, it gives one equation for both parts of the Green's equation with kinetic part satisfying diffusion equation and potential part satisfying rate equation. Solutions to these equations are well known,  a Gaussian spreading in \(\Delta \tau \) and an exponential function:
\begin{equation} \label{diff} 
G_{\rm diff}(x_b,x_a;\Delta\tau) = (4 \pi D \Delta\tau)^{-N/2}\exp[-(x_b-x_a)^2/4D\Delta\tau] \\
\end{equation}
and
\begin{equation} \label{branch} 
G_B(x_b,x_a;\Delta\tau) = \exp[-(\frac{1}{2}[V(x_a)+V(x_b)]-E_T) \Delta\tau],
\end{equation}
where the diffusion constant is \( D = \hbar^2/2m_e \) ( \( =1/2 \) in atomic units for the electron).

With these equations one can simulate random-walk-with-branching procedure to find the imaginary time evolution. Carrying out the simulation iteratively with short enough time step \( \Delta \tau \), large enough population of random walkers and adjusting the "trial energy" \( E_T \) to keep the simulation stationary will finally converge to the ground state wave function distribution of walkers and trial energy as the corresponding energy eigenvalue.

We should note, that Diffusion Monte Carlo method is generally used with trial wave functions \cite{DMCbook,Umrigar}, which makes DMC a significantly more powerful
tool than without, in its simple form, as introduced here.  Trial wave functions also enable studies of larger system sizes.  Also, the basic DMC is limited to simulation of ground states or wave functions without nodes \cite{FermionNodes,Tubman}, and furthermore, evaluation of other expectation values than total energy is not straightforward.

In this work, we will use the basic DMC without trial wave functions to make a fair comparison with our novel approach, and also, to be able to consider combination of these two Monte Carlo methods.

\goodbreak

\subsection{Real time propagation: RTPI}

For the real time dynamics of a quantum many-body system \( \psi(x,t) \) we define the Feynman path integral as
\begin{equation}    \label{kernel}  
	K(x_b,t_b;x_a,t_a) = \int_{x_a}^{x_b} \exp( {\rm i} S[x_b,x_a] ) \mathcal{D}x(t),
\end{equation}
where \(S[x_b,x_a] = \int_{t_a}^{t_b} L_x dt  \) is the action of the path \(x(t)\) from \((x_a,t_a)\) to \((x_b,t_b)\) and \( L_x \) is the corresponding Lagrangian \cite{feynman,feynman2}.  This is the kernel (or real time Green's function) of the propagation.

Now, the time evolution of the wave function \(\psi (x, t)\) (or probability amplitude), can  be written as
\begin{equation}   \label{propag}  
	\psi (x_b, t_b) = \int_a  K(x_b,t_b;x_a,t_a) \psi (x_a, t_a) {\rm d} x_a,
\end{equation}
where \(t_a < t_b\).  A more complete discussion about numerical time-dependent coherent PI solution for the full quantum dynamics is given elsewhere \cite{paper1}.

Now, we see the analogy of Eqs.~(\ref{SO5}) and (\ref{propag}), and the two propagators \( G \) and \( K \).  The latter of these is complex, bringing in the phase and interference of paths, an additional complication to numerical approaches, called "numerical sign problem" \cite{Kieu}.

\goodbreak

\subsection{Incoherent RTPI}

Here, we present the principle of incoherent real time propagation based on the numerical evaluation of \( \psi(x,t) \) in discretized time grid, stepwise with \( \Delta t \).

The real-time solution of the wave Eq. (\ref{SE1}) has the form
\begin{equation} \label{SO2} 
\psi(x,t) = \sum_{n=0}^{\infty} C_n \phi_n(x) \exp[-{\rm i}(E_n - E_T)t],
\end{equation}
analogously with the imaginary time solution, Eq. (\ref{SO1}).
By using the small angle approximation for short enough $\Delta t$ this can be written \cite{paper1} as
\begin{equation} \label{SO3} 
\psi(x,\Delta t) \approx \sum_{n=0}^{\infty} C_n \phi_n(x)\{1 - [( E_n - E_T)\Delta t]^2/2 -{\rm i}[ (E_n - E_T)\Delta t]\}.
\end{equation}
Now dropping off the imaginary part and keeping the real part of \( \psi \), only, the single step time evolution leads to
\begin{equation} \label{SO4} 
\psi_{\rm R}(x,\Delta t) = \sum_{n=0}^{\infty} C_n \phi_n(x) \{1 - [( E_n - E_T)\Delta t]^2/2\}.
\end{equation}
This incoherent propagation ignores the phase evolution by keeping the wave function real.

We see that the dominant term in the sum is the one, where \( \| E_n - E_T \| \) is least.  Therefore, the iterative propagation of \( \psi_{\rm R}(x,t) \) will converge to a real eigenstate $\phi_n$ with eigenenergy $E_n$ closest to \( E_T \), unless the initial \( \psi_{\rm R}(x,t) \) is orthogonal to it.  However, even in such case we can expect the numerical inaccuracies to generate a small seed of any eigenstate, and eventually, lead to the expected convergence.
In comparison of Eqs.~(\ref{SO1}) and (\ref{SO4}) we see that the imaginary time propagation can converge to the ground state of the system, only, while with the incoherent propagation $E_T$ can be chosen arbitrarily to find any non-degenerate eigenstate. This seems to be the most essential difference between RTPI and DMC.

In a graphical interpretation of Eq. (\ref{SO4}) the real wave function rotates in complex plane counterclockwise an angle \( ( E_n - E_T)\Delta t / \hbar \), and then, becomes projected back to the real axis, walker by walker.  The larger \( ( E_n - E_T) \Delta t \), the less of \( \phi_n(x) \) remains in the projection.  The hypothetical problem arising from the \( 2 \pi \) periodicity of the angle can be eliminated by changing or decreasing the time step.

\goodbreak

\subsection{Kernel and its approximations}

In general, the exact Kernel is rarely known and approximations are needed.  A usual approximation is sc.~"short time approximation" \cite{schulman, makri, paper1}
\begin{eqnarray}  \label{trottpropag}  
   K(x_b,x_a;  \Delta t) \approx \left[\frac{1}{2 \pi {\rm i}  \Delta t}\right]^{N/2} \exp \left[\frac{\rm i}{2 \Delta t} ( x_b - x_a )^2 - \frac{{\rm i} \Delta t}{2}(V(x_a) + V(x_b)) \right] ,
\end{eqnarray}
which becomes exact as \( \Delta t \rightarrow 0 \).  This is also called as Trotter kernel.

The product of Green's functions in Eqs. (\ref{diff}) and (\ref{branch}) is formally very similar to the propagator in Eq.~(\ref{trottpropag}), the exponential in the latter being complex causing oscillating wave behavior and numerical sign problem. This prevents interpretation of the Kernel straightforwardly as a probability.
We use the kernel of Eq.  (\ref{trottpropag}) for simulation of time evolution, as in Eq.  (\ref{propag}) in a short time step grid.  Regular grids may work, but as discussed earlier \cite{paper1}, Monte Carlo grids generate less artificial interferences.  We let the calculated real wave function \( \psi_{\rm R}(x) \) guide evolution of the grid in the role of walkers by using Metropolis Monte Carlo algorithm \cite{metropolis}.  Thus, in addition of the eigenenergy we obtain the wave function twice: calculated from the incoherent propagation, but also, represented by the walker distribution.

Wave oscillations in Eq. (\ref{trottpropag}) are strongest for large distances \( |x_a - x_b| \) and small time steps \( \Delta t \).  Therefore, the contributions of most paths (far from extrema of action) cancel efficiently by destructive interference, and in numerical calculations it is important not to waste time and efforts to those.  Instead, it is necessary to choose or weight the paths with most constructive interference, {\it i.e.}, those with large wave function amplitude.

Increase of the number of walkers (grid size) will also damp the artificial interference \cite{paper1} and there are other methods to make the kinetic term less oscillatory, {\it e.g.}, by using effective propagators 
\cite{filinov,makri4}.  Monte Carlo grids remove the interference arising from regularity, but bring in the roughness from randomness.  This may cause lack of sufficient destructive interference of paths where the walker density is low, {\it i.e.}, where the wave function decays to zero.  Smoothening of the roughness can be expected to help \cite{paper1}.

Our approach here is the following.  The initial wave function \( \psi (x_a, t_a) \) presented pointwise in a grid of walkers is "smoothened" to a "gaussianwise" presentation in the same grid by using Gaussian width parameter \( \epsilon \).
We make it by modifying the kinetic part of the propagator as
\begin{eqnarray}  \label{Gaussian}  
   \left[\frac{1}{2 \pi {\rm i}  \Delta t}\right]^{N/2} \exp \left[\frac{\rm i}{2 \Delta t} ( x_b - x_a )^2 \right]  \rightarrow
  \left[\frac{1}{2 \pi( {\rm i} \Delta t + \epsilon^2)}\right]^{N/2} \exp \left[\frac{\rm i}{2 (\Delta t - {\rm i} \epsilon^2)} ( x_b - x_a )^2 \right],
\end{eqnarray}
which converges back to the pointwise presentation as \( \epsilon \rightarrow 0 \).

The second part of kernel contributes to the numerical sign problem, if the potential is strongly variant along the paths, {\it i.e.}, if in Eq.~(\ref{trottpropag}) the change from \( V(x_a) \) to \( V(x_b) \) is far from linear or if \( (V(x_a)+V(x_b)) \) is locally variant for nearby paths.  To the latter, only increase of walker density helps, whereas to the former, averaging over the path or decrease of the time step \( \Delta t \) is needed.  Decrease of \( \Delta t \) leads essentially to decrease of \( |x_a - x_b| \) due to the "destructive kinetic interference".

Averaging over the path could be done with some pseudo-potential.  Such approximation could be tailored to include all the paths from \( x_a \) to \( x_b \) and remove the singularities, like the \( \Delta t \) dependent pair potential approximation widely used for the imaginary time propagation \cite{ceperley}.

Here, we adopt a straightforward single path average approximation
\begin{eqnarray}  \label{average}  
V_{\rm avg} = \frac{1}{x_b - x_a}  \int_{x_a}^{x_b} V(x) dx.
\end{eqnarray}
In one dimensional space this yields for the harmonic potential
\begin{eqnarray}  \label{harmonic}  
V_{\rm avg}^{\rm H} = \frac{\omega^2}{6} \left[ \frac{x_{1b}^3 - x_{1a}^3}{x_{1b}-x_{1a}} + \frac{x_{2b}^3 - x_{2a}^3}{x_{2b}-x_{2a}} \right]
\end{eqnarray}
and for the Coulomb potential
\begin{eqnarray}  \label{coulombic}  
V_{\rm avg}^{\rm C} = \frac {\ln (r_b / r_a )}{ r_b -  r_a },
\end{eqnarray}
where \( x_1 \) and \( x_2 \) are the particle coordinates, and \( r_a = x_{1a} - x_{2a} \) and \( r_b = x_{1b} - x_{2b} \) are the initial and final distances, as defined in the next section.

This result can also be acquired from a more sophisticated analysis and it is usually called semi-classical approximation \cite{ceperley}.

\goodbreak

\section{Hooke's Atom}

The problem of two electrons in a harmonic potential, Hooke's atom, has been investigated by several authors \cite{taut,qdot3,qdot4,qdot5,qdot6,qdot7,qdot8,qdot9,qdot10,qdot11,qdot12}.
There are even analytical solutions, but only for some specific parameters \cite{taut}.  There are also numerical solutions and some approximate approaches, but to the ground state energy and wave function, only \cite{sol1,sol2,sol3,sol4,sol5,sol6,sol7,sol8}.

\subsection{Separation of three dimensional problem}
Consider two electrons with Coulomb repulsion in a harmonic potential well.  With \( m_e = 1 \) the Hamiltonian of the system is
\begin{equation}\label{eq2H}
H(x_1,x_2) =-\frac{1}{2}\nabla_1^2-\frac{1}{2}\nabla_2^2+\frac{1}{2} \omega^2 x_1^2+\frac{1}{2} \omega^2 x_2^2+ \frac{1}{|x_1-x_2|},  
\end{equation}
where $x_1$ and $x_2$ are the three coordinates of two electrons.  The relative and center-of-mass (CM) motion of the electrons can now be separated by defining new three dimensional variables
\begin{eqnarray}\label{eq20}
r=x_1-x_2 {\rm \hskip 1cm and \hskip 1cm } R=\frac{x_1+x_2}{2}  
\end{eqnarray}
Thus, the Hamiltonian decouples as
\begin{equation}   \label{3Ddecoupled}                                
H(r,R) =-\frac{1}{2 \mu }\nabla_r^2+\frac{1}{2} \mu \omega^2 {r}^2 +  \frac{1}{|{r}|}-\frac{1}{2 M }\nabla_R^2+\frac{1}{2} M \omega^2 {R}^2 \equiv H_r + H_R,
\end{equation}
where $\mu= \frac{1}{2}$ and $M =2$ are the reduced and the total mass of electrons.  The wave function and total energy separates as \( \psi(r,R) = \phi(r) \Phi(R) \) and \(E = E_r + E_R\), respectively.

The CM motion is simple harmonic oscillation, which, of course, can further be separated into three one dimensional components. The relative motion of the two electrons is harmonic oscillation with the Coulomb repulsion as the perturbation.  This equation can be separated into radial and angular components similarly to the dynamics of the hydrogen atom.

With substitution \( \phi(r) = u(r)/r \) the radial equation of ground state takes the form
\begin{equation}\label{eq19-1}                                
[ -\frac{1}{2 \mu}\frac{{\rm d}^2}{{\rm d}r^2}+\frac{1}{2} \mu \omega^2 r^2 +\frac{1}{r} ] \ u(r) = E_r u(r).
\end{equation}
To find the exact solution we must solve a three step recurrence equation \cite{taut}, for which the solutions are restricted to some specific values of confinement parameters, only. For example, $\omega=\frac{1}{2}$ is one of the viable and the strongest confinement with ground state eigenenergy $E_r= \frac{10}{8}$ and wave function  \cite{taut}
\begin{equation}     \label{r-wf}                            
 u(r) = r \phi(r) = r\frac{ {\rm e}^{-r^2/8}\left(\frac{\left| r\right| }{2}+1\right)}{\sqrt{ \left(8+5 \sqrt{\pi }\right)}}.
\end{equation}

\subsection{Analytical solutions for one dimensional Hooke's atom}

Oseguera and Llano \cite{oseguera} have proven that for the attractive one-dimensional Coulomb potential, the singularity acts as an impenetrable barrier and the space becomes divided into two independent regions (space splitting effect). Therefore, the solutions for positive and negative values of the relative coordinates of two particle are completely independent. Due to the space spitting effect for one dimensional Coulomb potential, the wave function of the two particles should vanish where their relative coordinate becomes zero.

Because of this, the relative dynamics in one dimension is that of the radial part in three dimensions for the angular momentum quantum number \( \ell = 0 \), Eq.~(\ref{eq19-1}), \cite{taut}. With the definitions of $r$ and $R$ in Eq.~(\ref{eq20}), in one dimension
\begin{equation}                                             
\psi(r,R)=u(r)\Phi(R),
\end{equation}
where $u(r)$ is the relative motion wave function in one dimension Eqs.~(\ref{eq19-1}--\ref{r-wf}). It is related to the three dimensional relative motion wave function with zero angular momentum via $r \phi(r)=u(r)$.
In the one dimensional space the CM dynamics is simply that of one of the three \( R \)-components in Eq.~(\ref{3Ddecoupled}).  Thus, the ground state solution of \(H_R \Phi = E_{R} \Phi \) is
\begin{equation}   \label{R-energies}                                
\Phi(R) = \biggl( \frac{M \omega }{\pi } \biggr)^{1/4} {\rm e}^{- M \omega R^2 /2 } ,
\end{equation}
where \( M = 2 \) for electrons, and the corresponding energy is \( E_R = \frac{1}{2} \hbar \omega \).

We have chosen this as the test case for the numerical methods, below.  Thus, the ground state wave function \textit{\textbf{\( \psi(r,R) = u(r) \Phi(R) \)}} for one dimensional Hooke's atom with $\omega=\frac{1}{2}$  is given by Eqs.~(\ref{r-wf}) and (\ref{R-energies}), which yields the energy of
\begin{equation}                                         
E_0 = E_R + E_r = E_R + E_r^{\rm kin} + E_r^{\rm pot} = 0.25 + 0.28941 + 0.96058 = 0.25 + 1.25 = 1.5 
\end{equation}

\begin{figure}[t]                                       
\includegraphics[scale=0.47]{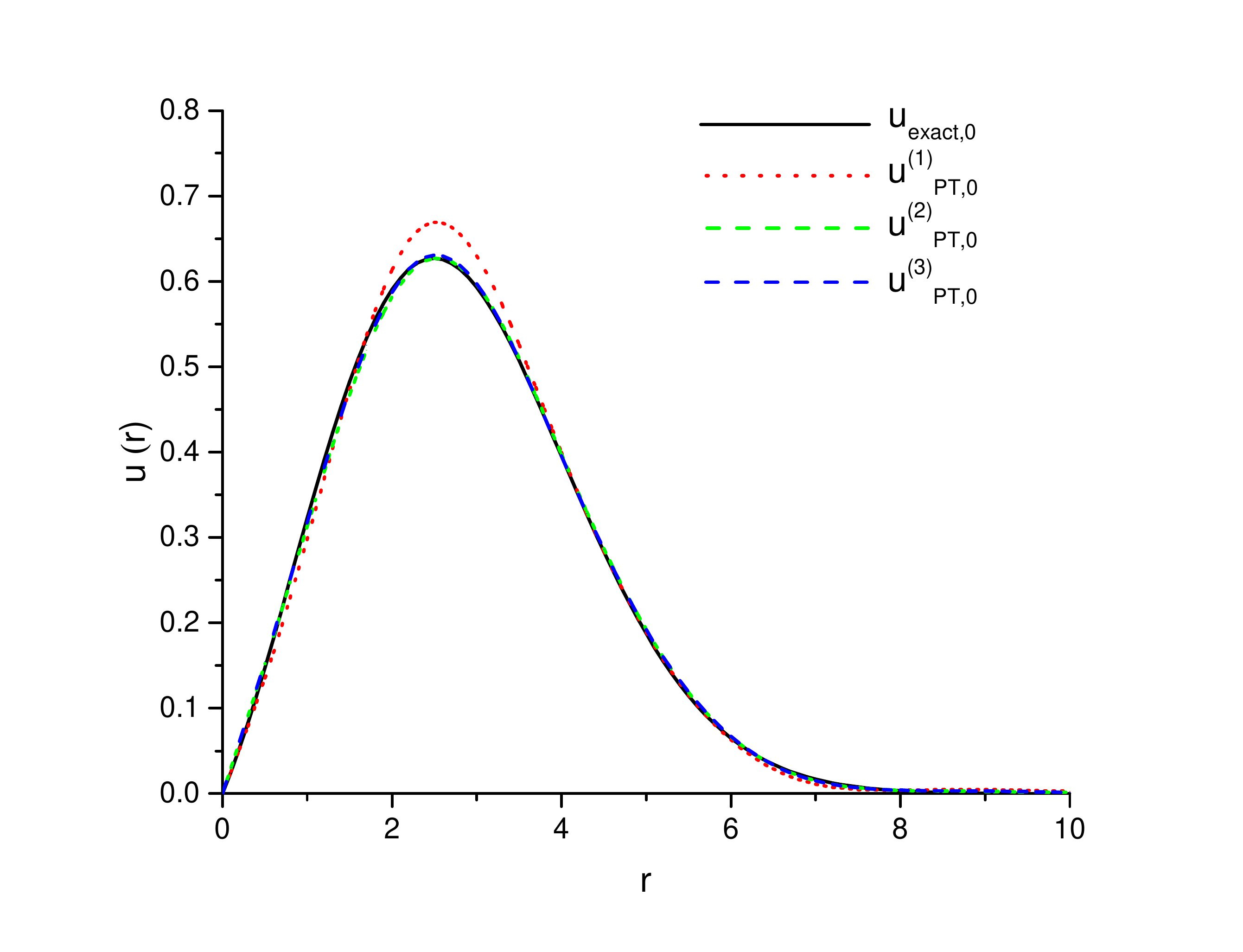}
\caption{Wave function of the ground state relative motion \( u_0(r) \) for $\omega = 1/2$, exact (solid line)\cite{taut} and from the three lowest order PT (red dotted, green dash-dotted and blue dashed lines) .}
\label{fig7}
\end{figure}

\begin{figure}[t]                                       
\includegraphics[scale=0.47]{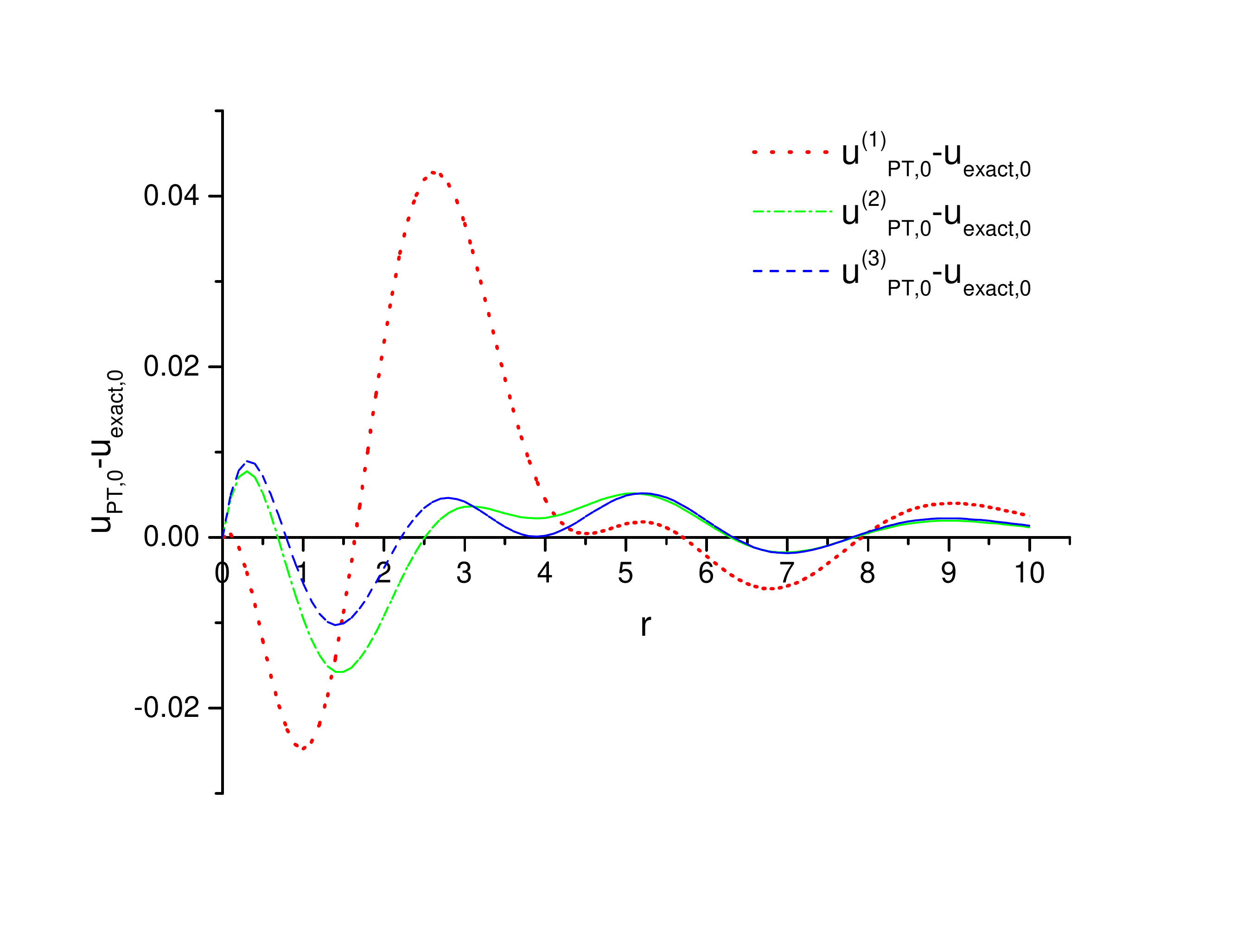}
\caption{The difference between the exact solution and
different order PT results are shown. The first, second and third order corrections are shown by the dots (red), dot-dash (green) and dash (blue) lines respectively.  The highest order PT has the best accuracy.}
\label{fig7b}
\end{figure}

The exact relative coordinate wave function is illustrated as a solid line in Fig.~\ref{fig7}, where it is also compared to those from PT. The differences between second order and third order perturbative solutions and exact solution are shown in Fig.~\ref{fig7b}.

\subsection{Solutions from perturbation theory}

Now, let us consider the Coulomb interaction as perturbation in Eq.~(\ref{eq19-1}).  Then, the unperturbed wave function and first order energy corrections are
\begin{eqnarray}\label{eq1}                                        
u^0_n (r)=\frac{\frac{1}{\sqrt[4]{\pi }} \sqrt{\xi } \exp \left(-\xi ^2 r^2 / 2\right) H_n(\xi  r)}{\sqrt{2^n n!}},\\
\delta E^1_n=2 e^2 \int_0^{\infty } \frac{u^{0*}_{2n+1}(r) u^0_{2n+1}(r)}{|r| } dr,\\
\xi=\sqrt{\frac{\mu \omega}{\hbar}},
\end{eqnarray}
where $H_n(r)$ are the Hermite Polynomials with \( n = 0, 1, 2, ...\)~.

To calculate the higher order energy corrections and wave function, we need matrix elements of the Coulomb potential. Generally, all of these can not be found in closed form, but for the ground state we can write
\begin{equation}                                          
V_{0,n}=\frac{e^2 \xi  \sqrt{2^{n+1} n!}}{n! \Gamma \left(1-\frac{n}{2}\right)}, \qquad\qquad n\geq 1.
\end{equation}
where $\Gamma$ is the Gamma function.

\begin{figure}[t]                                    
\includegraphics[scale=0.47]{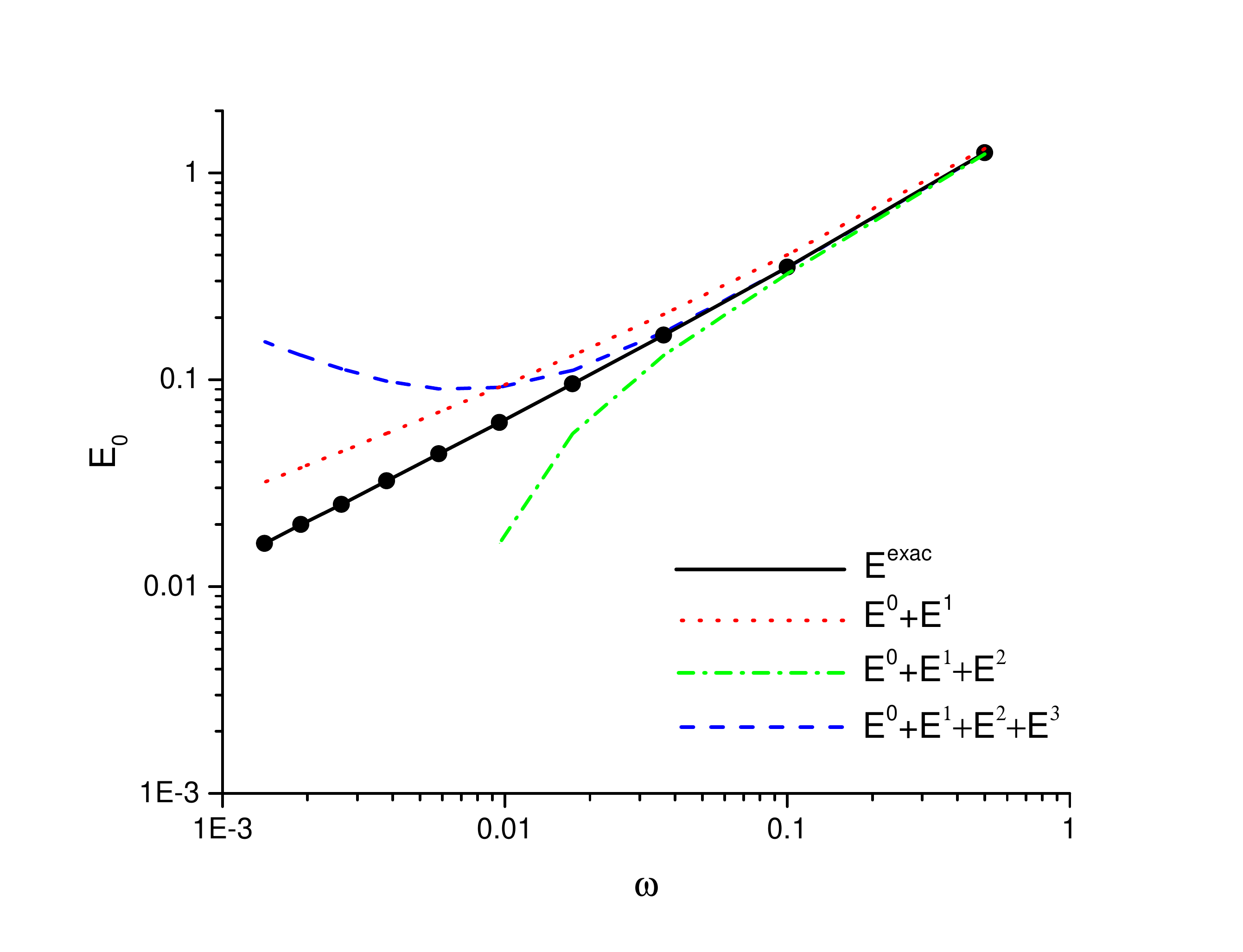}
\caption{Comparison of the ground state energies, exact \cite{taut} (black dots) and those from the PT (lines with the same notations as in Fig.~1).  The highest energy dot ($\omega = 1/2$) corresponds the wave functions in Fig. \ref{fig7}.}
\label{fig-pert}
\end{figure}

Then, up to the second order, the perturbative approximation of the ground state energy is
\begin{eqnarray}\label{energy}                             
E_0=\frac{3}{2} \hbar \omega + 2 e^2 \sqrt{\frac{\mu  \omega }{\pi \hbar }}+\sum _{n=1}^{\infty} \frac{|V_{0,2 n+1}|^2}{-2 n \omega  \hbar }.
\end{eqnarray}
Fig.~\ref{fig-pert} shows the ground state energy as a function of $\omega$. As the figure shows, for small values of $\omega$ the perturbative solutions do not converge to the exact values. We should notice that the PT is valid only as long as the condition $V_{nn} \ll |E_{n+1}-E_n|$ holds.

Average value of the kinetic energy can be used as a limit for validity of the perturbation method.  In the first order perturbative approximation, the average kinetic energy of relative motion for ground state, becomes negative for $\omega < 4 e^4 \mu / 9 \pi  \hbar ^3 \approx$ 0.07073 in au, and therefore, it can be chosen as the minimum acceptable frequency $ \omega_{\rm min} $ in perturbation method for ground state.

Clearly, the perturbation theory, Eq.~(\ref{energy}), works better for strong confinements, $\omega > 0.5$, where numerical approaches and other methods may become inaccurate.  For $\omega = 0.5$ the ground state energy for relative motion is 1.25 and perturbation theory yields 1.314, 1.238 and 1.248 for the 1st, 2nd and 3rd order, respectively, whereas 0.75 and 1.1830 is found by approximations for strong (ignoring e--e interaction) and weak (harmonic approximation) confinements \cite{taut}.

\goodbreak

\section{Monte Carlo simulations}

We assess performance and accuracy of numerical solutions by using the above defined one dimensional Hooke's atom with $k = \omega^2 = 1/4 $, {\it i.e.}, $ \omega = 0.5$, as the test bench.  First of all we consider our novel incoherent RTPI, as compared with diffusion Monte Carlo.  Finally, we introduce one more novel method by combining these two.

\subsection{Incoherent RTPI}

\begin{figure}

\includegraphics [trim=0cm 6cm 0cm 9cm, clip=true,scale=0.5] {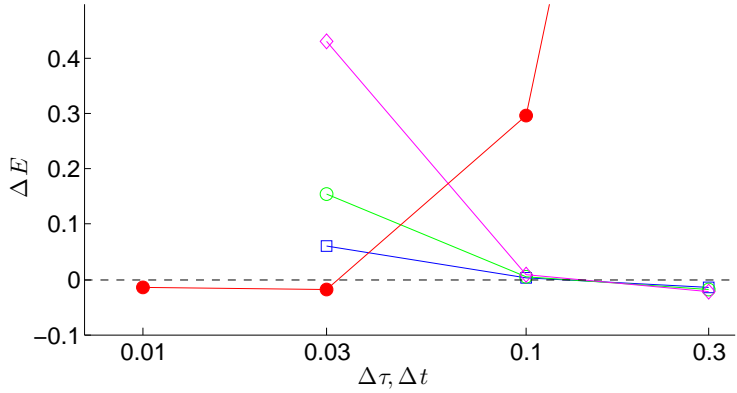}
\caption{Error in the calculated total energy $ \Delta E $ from its exact value $1.5$ as a function of time step in atomic units, $\Delta t$ (iRTPI) and $\tau$ (DMC). Symbols for RTPI are magenta open diamonds, green open circles and blue open squares for $N=$ 10k, 30k and 100k, respectively; and for DMC red full circles for 30k. Dashed line shows zero reference.}
\label{Energy}
\end{figure}

\begin{figure}

\includegraphics [trim=0cm 6cm 0cm 9cm, clip=true,scale=0.5]{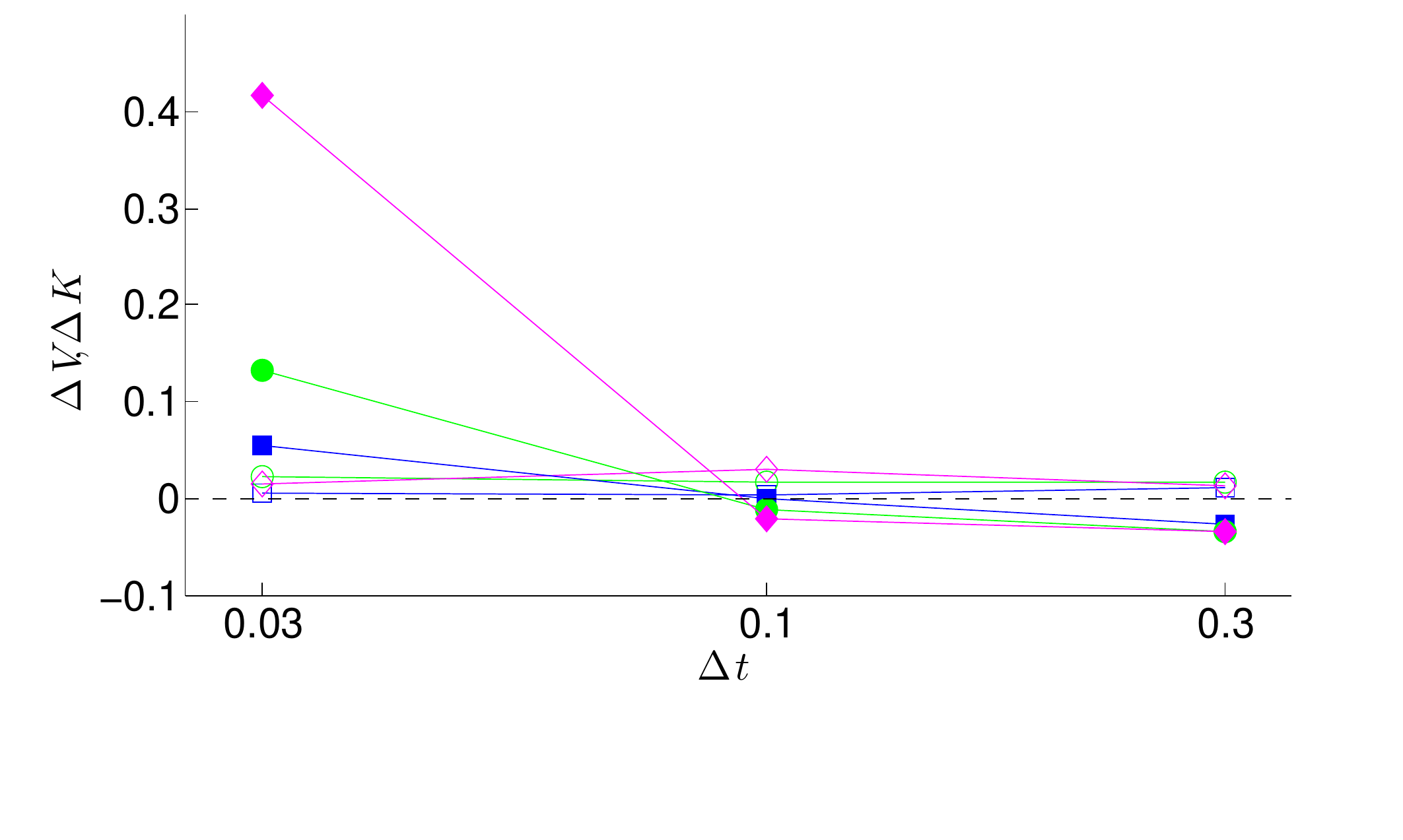}
\caption{Error in the calculated potential and kinetic energies, $ \Delta V $ and $ \Delta E - \Delta V $, from their exact values $1.0856...$ and $0.4144...$ respectively as a function of time step $\Delta t$. Open markers show the potential energy and filled markers the kinetic energy. Otherwise notations are the same as in Fig. \ref{Energy}. When comparing with Fig. \ref{Energy} we can see that most of the error comes from kinetic energy. }
\label{PotentialEnergy}
\end{figure}

To assess the performance of incoherent RTPI we first use constant width parameter $\epsilon^2 = 0.005$, see Eq. \ref{Gaussian}, and monitor the accuracy with respect to time step $\Delta t$ and grid size $N$. The data is collected in Table \ref{table1} and partly also shown in Figs.~\ref{Energy} and \ref{PotentialEnergy} in graphical form.

We see that the accuracy improves with increasing grid size, as expected. We find an optimal size for the time step, roughly at $\Delta t = 0.1$, for the modified Trotter kernel.  Shorter time steps strongly increase the error in the kinetic part of the propagator, whereas longer time steps increase that of the potential part, as discussed in sec.~2.4. Increasing time step also reduces the accuracy of Trotter kernel approximation. Shorter time steps can be used only with significantly larger grid size.

To allow estimation of convergence and statistical error of energies for practical calculations the standard deviations are given. It can be used to estimate the statistical accuracy (precision) of evaluated expectation values in form of standard error of mean, \( {\rm SEM} = \sigma / \sqrt{N} \), where \( N \) is the number of uncorrelated Monte Carlo steps.  Usually, \( 2 \times {\rm SEM} \) limits ( \( 95 \% \) ) are assumed as a statistical error estimate. For the present test cases evaluation of SEM is not relevant.

Next we keep the grid size and time step constant, $ N = 30 \times 10^3 $ and $ \Delta t = 0.1 $ and investigate the effect of the width parameter $\epsilon^2$. The data is collected in Table  \ref{table2}. Obviously, the larger the grid size, the smaller $ \epsilon $ can be used, increasing the accuracy. However, $ \epsilon $ at least of the size of the average distance of walkers "smoothens" the wave function also increasing accuracy.  On the other hand, the last line of Table \ref{table2} proves that $ \epsilon \approx 0.3 $ is still applicable, though the value $ \epsilon \approx 0.07. $ ($ \epsilon^2 = 0.005 $) was found the best with $ N = 30 \times 10^3$.

\begin{table} 

\caption {Accuracy and distribution of energetics in incoherent RTPI simulations of the ground state. N is the number of walkers (\({\rm k}=10^3\)), \( \Delta t \) the time step,  \( \Delta E \)  deviation of expectation values from the exact value \( 1.5 \), \( \Delta V \) deviation of expectation value from the exact value \( 1.0856... \), \( \sigma \) standard deviation in 20 blocks of data with 50 iterations in block and \( \epsilon^2 \) the "gaussian width of walkers" discussed in section 2.2. (All quantities in atomic units)}

\vskip 0.2cm \centering
\begin{tabular} {c c c c c c c}
\hline \hline

$ $N$ $ & $\Delta t$ & $\Delta E $ & $\sigma_E $ & $ \Delta V $ & $\sigma_V $ & $ \epsilon^2 $ \\ [0.5ex]
\hline
  $ 100{\rm k} $& $ 0.3 $  &   $ -0.0152 $ & $ 0.0008 $ & $ 0.0109$ & $0.0017 $ & $0.005 $\\
  $ 100{\rm k} $& $ 0.1 $  &   $ 0.0032 $ & $ 0.0014 $ & $ 0.0039$ & $ 0.0021 $ & $0.005 $\\
  $ 100{\rm k} $& $ 0.03 $ &   $ 0.0601 $ & $ 0.0135 $ & $ 0.0054$ & $ 0.0014 $ & $0.005 $\\
  $ 30{\rm k} $& $ 0.3 $   &   $ -0.0185 $ & $ 0.0016 $ & $ 0.0161$ & $0.0042 $ & $0.005 $\\
  $ 30{\rm k} $& $ 0.1 $   &   $ 0.0046 $ & $ 0.0058 $ & $ 0.0172$ & $ 0.0111 $& $0.005 $ \\
  $ 30{\rm k} $& $ 0.03 $  &   $ 0.1544 $ & $ 0.0333 $ & $ 0.0221$ & $ 0.0247 $& $0.005 $ \\
  $ 10{\rm k} $& $ 0.3 $   &   $-0.0220 $ & $0.0030 $ & $ 0.0126 $ & $ 0.0062$ & $0.005 $ \\
  $ 10{\rm k} $& $ 0.1 $   &   $ 0.0077 $ & $ 0.0123 $ & $ 0.0296$ & $ 0.0505 $& $0.005 $ \\
  $ 10{\rm k} $& $ 0.03 $  &   $ 0.4324 $ & $ 0.0653 $ & $ 0.0154$ & $ 0.0045 $& $0.005 $ \\
\hline
\end{tabular}
\label{table1}
\end{table}

\begin{table} 

\caption {Effect of the gaussian width of walkers. Notations are the same as in Table \ref{table1}.}

\vskip 0.2cm \centering
\begin{tabular} {c c c c c c c}
\hline \hline

$ $N$ $ & $\Delta t$ & $\Delta E $ & $\sigma_E $ & $ \Delta V $ & $\sigma_V $ & $ \epsilon^2 $ \\ [0.5ex]
\hline
  $ 30k $& $ 0.1 $   &   $- 0.0235 $ & $ 0.0068 $ & $ 0.0510$ & $ 0.0168 $& $0 $ \\
  $ 30k $& $ 0.1 $   &   $ 0.0025 $ & $ 0.0037 $ & $ 0.0146$ & $ 0.0112 $& $0.005 $ \\
  $ 30k $& $ 0.1 $   & $-0.0118   $ & $0.0024 $ & $ 0.0848 $ & $ 0.0194$ & $0.1 $ \\

\hline
\end{tabular}
\label{table2}
\end{table}

\begin{figure}

\includegraphics [trim=0cm 6cm 0cm 7cm, clip=true,scale=0.5]{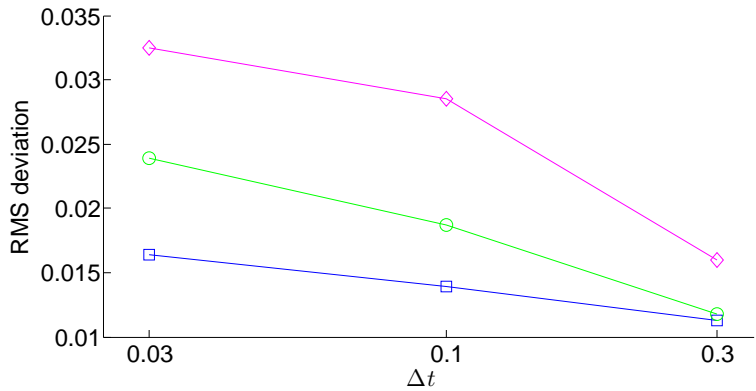}
\caption{RMS deviation of the calculated and analytical wave functions as a function of time step $\Delta t$. Notations are the same as in Fig.~\ref{Energy}.}
\label{RMS}
\end{figure}

In Fig.~\ref{RMS} the difference between calculated wave function and the exact one is shown in terms of the root-mean-square deviation as a function of time step length.  This data have the same trends as the error in total energy in Fig. \ref{Energy}.  

\begin{figure}

\includegraphics[trim=2cm 8cm 3cm 8cm, clip=true,scale=0.5]{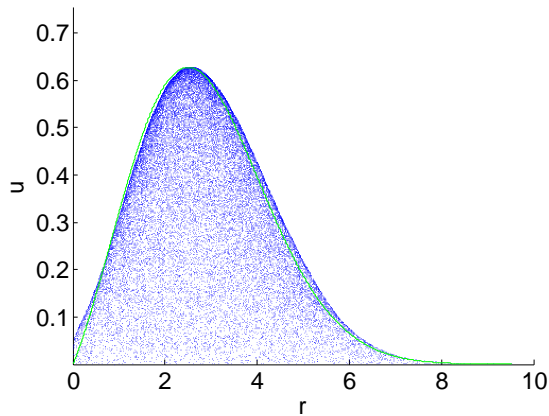}
\caption{Snapshot of the simulation for relative motion wave function with N = 30k, \(\Delta t \) = 0.1 and $\epsilon^2 = 0.1$.  The blue dots show walkers projected onto the plane $ (r, u(r)) $.  Green line shows the corresponding exact wave function $u(r)$, Eq.~(\ref{r-wf}) for $ R_{CM} = 0$. The maximum amplitude of walkers is scaled to match the maximum of the analytical maximum and $r$ is in and atomic units.}
\label{epsilon01}
\end{figure}

For more detailed analysis of the wave function we will consider the relative motion and CM motion separately.  First, the Fig. \ref{epsilon01} shows a plot of all walkers (a snapshot of Monte Carlo grid points) as red dots at the walker coordinates $ (r, u(r)) $, {\it i.e.}, the relative coordinate and real amplitude.  The green line shows the maximum amplitude given by Eq.~(\ref{r-wf}).  This is the simulation case of the last line of Table \ref{table2}.  We see that the wave function match is pretty good, but the energetics is not.

\begin{figure}

\includegraphics[trim=2cm 8cm 3cm 8cm, clip=true,scale=0.55]{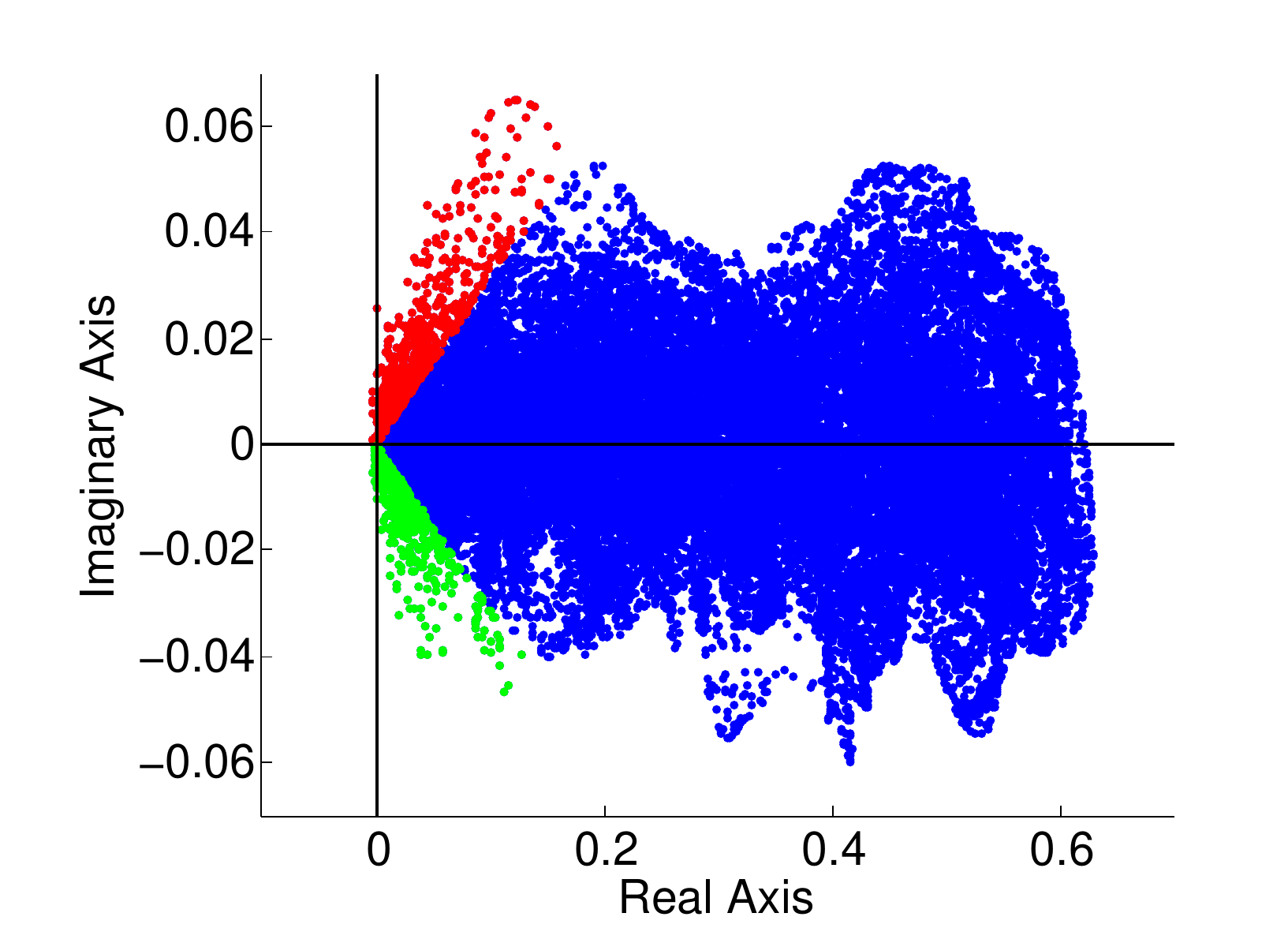}
\caption{The complex wave function phase evolution of the real ground state in one time step ($\Delta \varphi = 0$ expected.).  Note the different scaling of axes. Color coding: red for $\Delta \varphi > 0.3$ and green for $\Delta \varphi < -0.3$. These walkers are more than 20\% off from the known expectation value of energy. N = 30000, \(\Delta t \) = 0.1 and $\epsilon^2 = 0.005$. }
\label{Phase1}
\end{figure}

Another snapshot of walkers is shown in Fig. \ref{Phase1}.  This plot presents the complex amplitude in complex plane after one time step starting from the real wave function, Eq.~(\ref{SO4}), {\it i.e.}, starting from all the walkers lying in the real axis according to the amplitude.  With the walkers perfectly representing the stationary state one expects no other changes than rotation of all walkers around origin corresponding to the phase $\varphi \propto (E_0-E_T) / \Delta t $ and retaining their absolute value (modulus).  Fig. \ref{Phase1} shows the case, where $ E_T = E_0 $, and thus, $\varphi = 0 $ is expected.  Therefore, deviations from the real axis present numerical or statistical error emerging from the random nature of Monte Carlo grid in the figure. 
Thus, the imaginary components of amplitude are zero in average and large deviations from the average, shown in red and green, point out the walkers with largest error.  Note the different scaling of real and imaginary axes.

\begin{figure}

\includegraphics[trim=2cm 8cm 3cm 8cm, clip=true,scale=0.55]{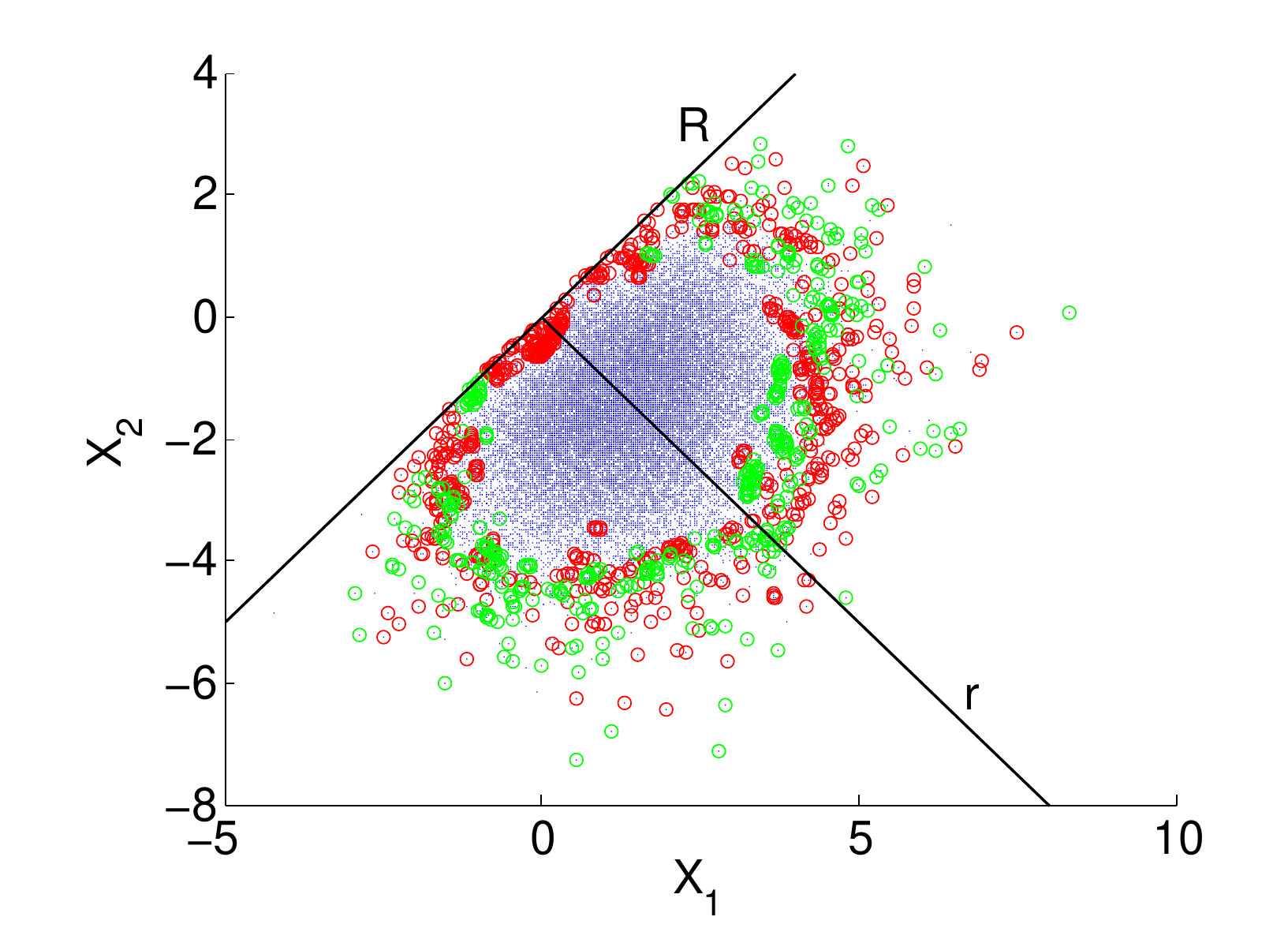}
\caption{Plot of the same walkers as in Fig.~\ref{Phase1}, but now in a plane of coordinates of electrons  $ (x_1, x_2) $ in atomic units. The color coding is the same, but size of blue and other walkers is smaller and larger, respectively.  The CM and relative coordinate axes are also shown.}
\label{Phase2}
\end{figure}

The same walkers as in Fig.~\ref{Phase1} are shown in Fig.~\ref{Phase2}, but now in two dimensional cartesian coordinate system of $ (x_1, x_2) $, the coordinates of the two electrons in one dimension.  Now, we see that the walkers showing the most erroneous phase in Fig.\ref{Phase1} are those in the region with the lowest density, where the wave function decays to zero.  Obviously, the sparse walker density is not able to describe the amplitude smoothly, enough.

The sparse grid problem is expected but can not be avoided, in case the walker density represents the wave function amplitude in the region of almost vanishing wave function.  Increase of the walker width parameter $\epsilon^2$ helps until it starts to adversely round the shape of the wave function in other regions, see Fig.~\ref{epsilon01}.

\goodbreak

\begin{figure}[t]                                        

\includegraphics[trim=2cm 8cm 2cm 8cm, clip=true,scale=0.45]{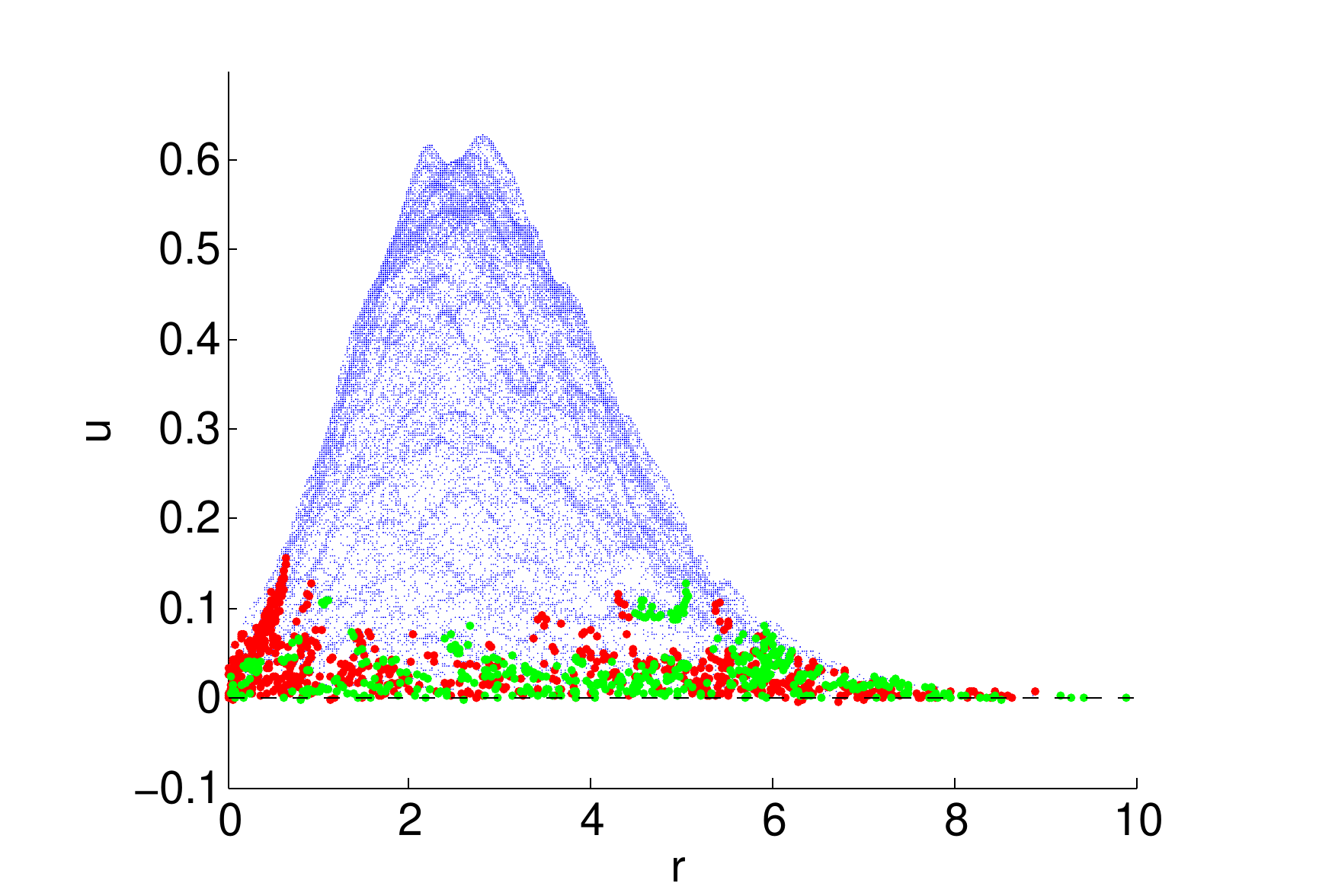}
\caption{Snapshot of the calculated wave function in the relative motion coordinates with  N = 30000, \(\Delta t \) = 0.1 and $\epsilon^2 = 0.005$. The walkers and the color coding are the same as in the Figs.~\ref{Phase1}--\ref{Phase2}. Units are the same as in \ref{epsilon01}}
\label{Phase3}
\end{figure}

\begin{figure}[h]                                        

\includegraphics[trim=2cm 8cm 2cm 8cm, clip=true,scale=0.45]{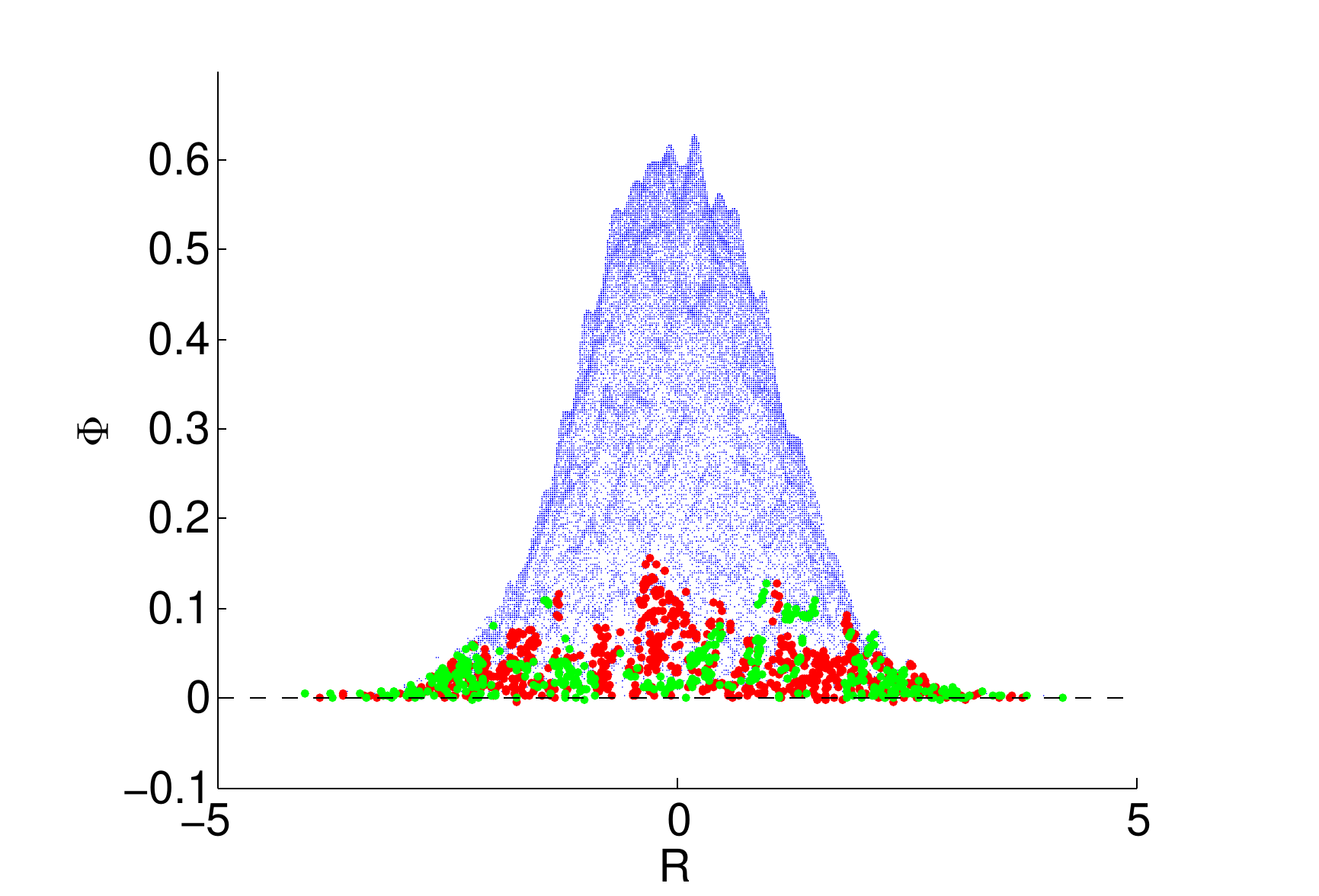}
\caption{Snapshot of the calculated wave function in the center of mass coordinates with  N = 30000, \(\Delta t \) = 0.1 and $\epsilon^2 = 0.005$. The walkers and the color coding are the same as in the Figs.~\ref{Phase1}--\ref{Phase3}. Units are the same as in \ref{epsilon01}}
\label{Phase4}
\end{figure}

\vskip 2 cm

In Figs. \ref{Phase3} and \ref{Phase4} two different projections of again the same set of walkers is shown.  The real amplitude is presented as a function of the relative and CM coordinate, {\it i.e.}, walkers in planes $ (r, u(r)) $ and $ (R, u(r)) $, respectively.  Thus, Fig.\ref{Phase3} is the same projection as that in Fig. \ref{epsilon01}, but now from a simulation with better numerical parameters as judged by the accuracy of energetics in Table \ref{table1}.

Now, we see strong fluctuations in amplitude with the RMS deviation from the exact wave function given in Fig. \ref{RMS}.  Cumulative averaging over the Monte Carlo simulation, however, smoothens these fluctuations away.

\goodbreak

\subsection{Simple DMC}

Diffusion Monte Carlo in its most efficient form includes the use of trial wave functions. Here we introduced only the simple DMC approach in order to allow straightforward comparison with RTPI. Moreover, simple DMC and RTPI can be combined to a novel method with new practical properties that simplify the calculation of various observables, not just the total energy. This will be discussed in the next subsection in more detail. 

Since Hooke's atom does not involve attractive singular Coulomb potentials, only \( + \frac{1}{r} \), the Trotter break-up is valid and the branching term in Eq.~(\ref{branch}) does not diverge. Therefore, sampling without a trial wave function can be expected to give accurate results with sufficiently small imaginary time time step and large enough number of walkers.

The data in Table \ref{table3} shows that the total energy converges to its exact value as the imaginary time step \( \tau \rightarrow 0 \).
By comparing the data in Table.~\ref{table1} we see that with optimal parameters and the same number of walkers $N$, RTPI gives similar accuracy as simple DMC, if only the number of Monte Carlo steps matters.

However, RTPI is computationally much more demanding. This stems from the fact, that for each MC step in DMC algorithm, only $N$ moves of walkers guided by the potential function is needed, but with the present incoherent RTPI we need to calculate $N \times N$ real time propagations to evaluate the guiding distribution before moving the walkers. 

\begin{table} [t]                                       

\caption {Accuracy and distribution of energetics in DMC simulations of the ground state. Number of walkers is $ N = 30k $, \( \tau \) is the imaginary time step,  \( \Delta E \)  deviation of the expectation value from its exact value  \( 1.5 \) and \( \sigma \) is the standard deviation of 20 blocks of data.  Each block consists of 50 iterations.  A new energy estimate was calculated after each block.
}

\vskip 0.2cm \centering
\begin{tabular} {c c c}
\hline \hline

$\tau$ & $\Delta E $ & $\sigma $ \\ [0.5ex]
\hline
  $ 1 $   &   $-0.0526 $ & $0.0008$ \\
  $ 0.3 $ & $-0.0197$ & $0.0016$ \\
  $ 0.1$      &   $ -0.0096 $ & $0.0034$ \\
  $ 0.03$  & $ -0.0063 $ & $0.0049$ \\
  $ 0.01$      &   $ -0.0041 $ & $0.0085$ \\
  $ 0.001$      &   $0.0137 $ & $0.0235$ \\
\hline
\end{tabular}
\label{table3}
\end{table}

\subsection{Combination of DMC and RTPI}
The simple DMC algorithm samples the nodeless ground state distribution, which makes it cumbersome and inefficient to evaluate expectation values for observables other than the total energy, even as simple as the potential energy.  This is due to the availability of the wave function in form of walker distribution, only.  Thus, in evaluation of the expectation  value matrix elements the walker distribution implicitly contributes as the bra wave function, but ket is not available.

For direct sampling of matrix elements another representation of the wave function, the ket vector, is needed.  For this purpose the incoherent RTPI on DMC walkers can be used.  Then, also the overlap integral and potential energy becomes straightforward to sample for
\begin{eqnarray}  \label{expectation}  
\langle V \rangle = \frac{ \langle \psi \mid V \mid \psi \rangle}{ \langle \psi \mid \psi \rangle}.
\end{eqnarray}
Similarly, expectation values for any local multiplicative operators become available.
Also, another total energy estimate is obtained from sampling the phase evolution in real time.

The incoherent RTPI can be used to evaluate, not only the ground state, but also the excited states with the positive and negative amplitudes, and thus, it provides means for locating the nodal surfaces.  This suggests combining the two approaches for evaluation of excited states, or in general, states with nodes in the spirit of released nodes idea; RTPI would be used in finding the nodes and evaluating another wave function, as well as another total energy, while DMC is used to sample the paths for RTPI.  This has potential for increasing the usefulness and capabilities of the simple DMC approach.

For practical use the number of RTPI steps can be kept much less than the DMC steps.  In case nodal information is not needed, only one RTPI step at the end may be sufficient.  Evaluation of statistical precision calls for more than that, of course.

In Table \ref{DMCwRTPI} we show the data evaluated with the combined approach.  The underlying DMC has been run with \( \tau = 0.01 \), see Table \ref{table3} and RTPI on top of that with \( \Delta t = 0.1 \) with the optimal choice of other parameters, see above.  RTPI step has been run once for every other block of 50 DMC steps.

\begin{table} 

\caption {Incoherent RTPI combined with DMC.  The walker distribution $ N = 30k $ is sampled by DMC with ($\tau = 0.01$) and RTPI step length is $\Delta t = 0.1$.  Evaluated expectation values and DMC total energy are given with their standard deviations.  Notations are the same as in the previous Tables. DMC is calculated from 20 blocks of data with 50 iterations per block. RTPI step is run once for every other block.}

\vskip 0.2cm \centering
\begin{tabular} {c c c c c c}
\hline \hline

& $\Delta E $ & $\sigma_E $ & $ \Delta V $ & $\sigma_V $ & $ \epsilon^2 $ \\ [0.5ex]
\hline
RTPI & $ 0.0033 $& $ 0.0060 $   &   $ 0.0022 $ & $ 0.0039 $ & $ 0.005$\\
\hline
DMC & $-0.0041 $& $ 0.0085 $   &   \\
\hline
\end{tabular}
\label{DMCwRTPI}
\end{table}

\begin{figure}[t]                                       

\includegraphics[trim=2cm 7.5cm 2cm 8.5cm, clip=true,scale=0.43]{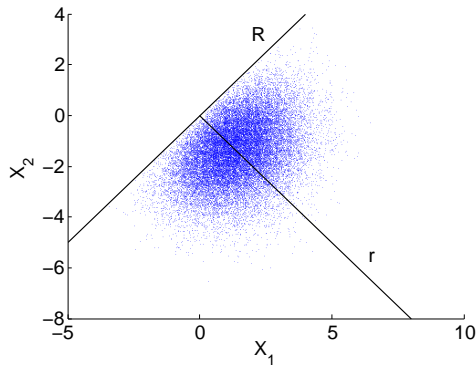}
\caption{DMC simulation snapshot of walker distribution in a plane of coordinates of electrons  $ (x_1, x_2) $ and separated coordinates $ (r, R) $, in atomic units.  Parameters \( \tau = 0.001 \) and $ N = 30000 $ were used.}
\label{DMC1}
\end{figure}

We find that DMC sampling of walkers from the distribution derived from the potential function leads to smoother spatial distribution than that of guided by the wave function amplitude from RTPI.  This can be clearly seen by comparing the distributions in Figs.~\ref{Phase2} and \ref{DMC1}, and also, the amplitudes in Figs.~10 and 13. This also yields better energetics which can be seen by comparing the values from Table.~\ref{table1} and \ref{DMCwRTPI}.  In the latter one there are less stray walkers at very low density region.  We assume that the reason for this is in the different nature of the guiding distribution: for DMC it is stable well-defined potential, while for the Metropolis algorithm in RTPI it is the calculated amplitude presented in the Monte Carlo grid.  This problem is always present at the region of low walker density.

Thus, if more stability is needed and larger number of walkers becomes too expensive, it may be necessary to use cumulative distribution of the amplitude from several previous RTPI steps.  According to our preliminary testing, same type of problem may arise in locating the nodal surfaces accurately enough.  Use of the cumulative distributions calls for numerical algorithms for efficient interpolation and updating the collected data.

\begin{figure}[h]                                       

\includegraphics[trim=2cm 7.5cm 2cm 8.5cm, clip=true,scale=0.43]{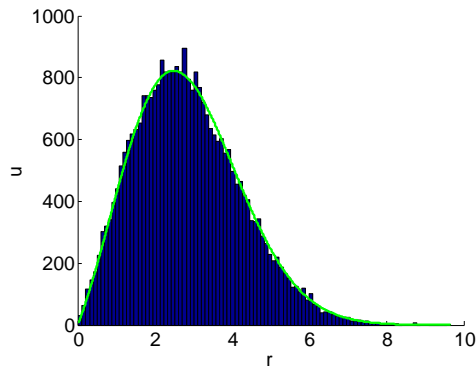}
\caption{Snapshot histogram of walker distribution in DMC simulation with \( \tau = 0.001 \) and N = 30000. \( \mid x_b - x_a \mid \) projection. Green line is the analytical solution fitted to the data. $u$ is the number of walkers in each bin and $r$ is in atomic units. }
\label{DMC2}
\end{figure}

Based on the experience, so far, the combination of the simple Diffusion Monte Carlo and incoherent Real Time Path Integral methods form a novel approach for electronic structure calculations that is capable of extending usefulness of the former. However, its significance remains to be tested in practice.

\section{Conclusions}

We have introduced a new approach based on the incoherent real
time path integrals (iRTPI) for ground and excited state electronic
structure calculations.  It includes correlations between electrons
exactly, within the numerical accuracy, which can be made better than the systematic error from the kernel simply by using Monte Carlo technique.  Here, we use Hooke's atom, a
two-electron system with very strong correlation, as our test case,
which we solve with both iRTPI and diffusion Monte Carlo
(DMC) for comparison. The high accuracy and stability of iRTPI is
demonstrated, and the improved Trotter kernel is shown to be useful
with large enough number of Monte Carlo walkers. In
addition, useful numerical parameters for the present case have been
determined for stable and self-consistent simulations. 

In its present form the computational cost of iRTPI is significantly
higher than that of DMC.  However, one of the advantages of iRTPI is that
it provides one with the wave function explicitly, and thus, the evaluation of local multiplicative expectation values becomes straightforward. Moreover, it is also capable of
locating excited states, and thus, the related nodal surfaces, the technical details of which were not considered here.  In addition,
incoherent dynamics can be turned to coherent dynamics, in case quantum dynamics is relevant.

We also showed that another novel approach obtained by combining the
iRTPI and DMC methods allows a more straightforward means for
evaluation of various observables within the robust framework of DMC. Due the
capability of iRTPI for locating the nodal surfaces, it will be
interesting to further test this combination method in a released node
fashion of DMC. This would mean a trial wave function free DMC also
for fermions.

Perturbation theory was shown to be useful for analytical solutions in case of strong confinements, which may become more challenging for numerical methods and available approximate solutions.  On the other hand, for weak confinements, {\it e.g.} in quantum dots, the presented numerical iRTPI method is expected to be robust.

\section*{Acknowledgements}

For computational resources we like to thank the Techila Technologies facilities and TCSC at Tampere University of Technology, and also, services of the Finnish IT Center for Science (CSC).

\end{document}